\documentclass[sigconf]{acmart}
\AtBeginDocument{%
  }

\setcopyright{acmlicensed}
\copyrightyear{2018}
\acmYear{2018}
\acmDOI{XXXXXXX.XXXXXXX}
\acmConference[Conference acronym 'XX]{Make sure to enter the correct
  conference title from your rights confirmation email}{June 03--05,
  2018}{Woodstock, NY}
\acmISBN{978-1-4503-XXXX-X/2018/06}

\usepackage{multirow}
\usepackage{booktabs}
\usepackage{graphicx}
\usepackage{arydshln}
\usepackage{bbm}
\usepackage{url}
\usepackage{tikz}
\usetikzlibrary{shapes.geometric, arrows.meta, positioning, calc}
\usepackage{caption}
\usepackage{amsmath}
\begin{document}

\title{LaDA-Band: Language Diffusion Models for Vocal-to-Accompaniment Generation}

\author{Qi Wang}
\authornote{Both authors contributed equally to this research.}
\affiliation{%
  \institution{State Key Lab of AI Safety, Institute of Computing Technology, Chinese Academy of Sciences (CAS) \\ University of Chinese Academy of Sciences}
    \city{Beijing}
  \country{China}}
\email{wangqi245@mails.ucas.ac.cn}

\author{Zhexu Shen}
\authornotemark[1]
\affiliation{%
  \institution{Lyra Lab, Tencent Music Entertainment}
  \city{Shenzhen}
  \country{China}}
\email{zhexushen@tencent.com}

\author{Meng Chen}
\affiliation{%
  \institution{Lyra Lab, Tencent Music Entertainment}
  \city{Shenzhen}
  \country{China}}
\email{yihuanchen@tencent.com}

\author{Guoxin Yu}
\authornote{Corresponding Author}
\affiliation{%
  \institution{Pengcheng Laboratory}
  \city{Shenzhen}
  \country{China}}
\email{yugx@pcl.ac.cn}

\author{Chaoxu Pang}
\affiliation{%
  \institution{State Key Lab of AI Safety, Institute of Computing Technology, Chinese Academy of Sciences (CAS) \\ University of Chinese Academy of Sciences}
    \city{Beijing}
  \country{China}}
\email{pangchaoxu21b@ict.ac.cn}

\author{Weifeng Zhao}
\affiliation{%
  \institution{Lyra Lab, Tencent Music Entertainment}
  \city{Shenzhen}
  \country{China}}
\email{ethanzhao@tencent.com}

\author{Wenjiang Zhou}
\affiliation{%
  \institution{Lyra Lab, Tencent Music Entertainment}
  \city{Shenzhen}
  \country{China}}
\email{alanzhou@tencent.com}
\renewcommand{\shortauthors}{Trovato et al.}

\begin{abstract}
Vocal-to-accompaniment (V2A) generation, which aims to transform a raw vocal recording into a fully arranged accompaniment, inherently requires jointly addressing an accompaniment trilemma: preserving acoustic authenticity, maintaining global coherence with the vocal track, and producing dynamic orchestration across a full song. Existing open-source approaches typically make compromises among these goals. Continuous-latent generation models can capture long musical spans but often struggle to preserve fine-grained acoustic detail. In contrast, discrete autoregressive models retain local fidelity but suffer from unidirectional generation and error accumulation in extended contexts. We present LaDA-Band, an end-to-end framework that introduces Discrete Masked Diffusion to the V2A task. Our approach formulates V2A generation as Discrete Masked Diffusion, i.e., a global, non-autoregressive denoising formulation that combines the representational advantages of discrete audio codec tokens with full-sequence bidirectional context modeling. This design improves long-range structural consistency and temporal synchronization while preserving crisp acoustic details. Built on this formulation, LaDA-Band further introduces a dual-track prefix-conditioning architecture, an auxiliary replaced-token detection objective for weakly anchored accompaniment regions, and a two-stage progressive curriculum to scale Discrete Masked Diffusion to full-song vocal-to-accompaniment generation. Extensive experiments on both academic and real-world benchmarks show that LaDA-Band consistently improves acoustic authenticity, global coherence, and dynamic orchestration over existing baselines, while maintaining strong performance even without auxiliary reference audio. Codes and  audio samples are available at \url{https://github.com/Duoluoluos/LaDA-Band} .
\end{abstract}



\received{20 February 2007}
\received[revised]{12 March 2009}
\received[accepted]{5 June 2009}

\maketitle

\section{Introduction}

\begin{figure}[t]
  \centering
  \includegraphics[width=0.9\linewidth]{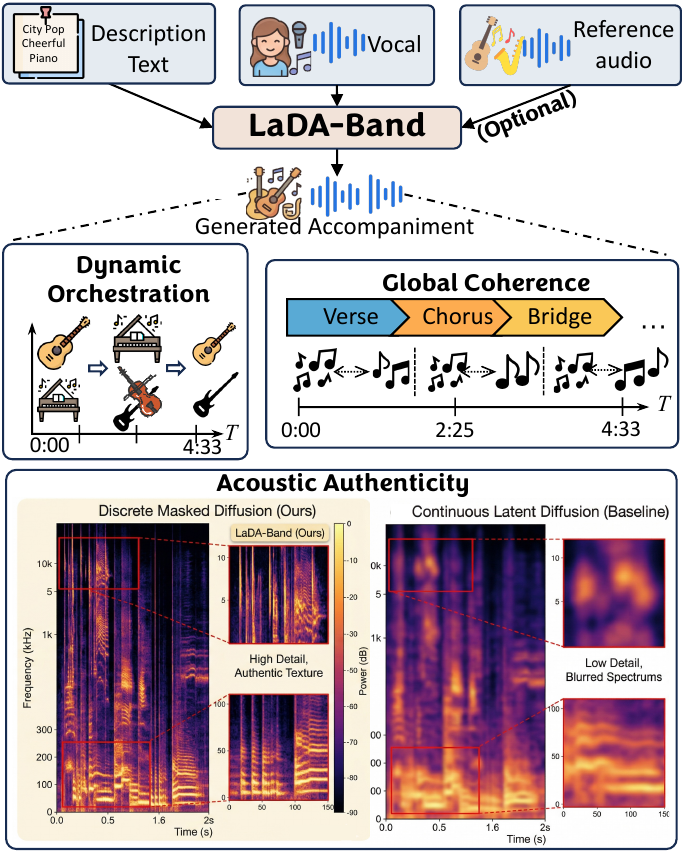} 
\caption{LaDA-Band enables true zero-shot generation (i.e., generating full arrangements from only vocals and text, without needing reference audio).}
  \label{fig:teaser}
  \vspace{-2mm}
\end{figure}

Music creation often begins with a simple musical idea, such as a hummed melody or a recorded dry vocal. Converting such raw input into a fully produced song requires accompaniment generation, which we study as the task of vocal-to-accompaniment (V2A) generation. Unlike general text-to-music synthesis~\cite{copet2023simple, yuan2025yue}, zero-shot V2A generation must simultaneously satisfy three key requirements, which together form the Accompaniment Trilemma:
1) \textit{\textbf{Acoustic Authenticity}}: generating accompaniment with fine-grained acoustic details and natural timbral quality, while avoiding perceptible artifacts.
2) \textit{\textbf{Dynamic Orchestration}}: arranging the accompaniment so that it changes appropriately over the course of the song.
3) \textit{\textbf{Global Coherence}}: maintaining harmonious alignment with the vocal track while preserving full-song structure.
As shown in Figure~\ref{fig:teaser}, a practical zero-shot V2A system should generalize directly from the raw vocal input, without depending on auxiliary reference accompaniments or regenerated vocals at inference time.

Despite the rapid evolution of generation architectures, existing open-source V2A systems still struggle to satisfy this trilemma, largely due to limitations in both generation formulation and representation space. First, \textit{discrete autoregressive} (\textit{Disc. AR}) models, represented by SongEditor~\cite{yang2025songeditor}, are inherently difficult to scale to full-song generation. Although these models can produce locally coherent short segments, their token-by-token decoding incurs prohibitively high memory and computational costs as sequence length grows. More importantly, autoregressive decoding conditions each new token on previously generated outputs, causing errors to accumulate over time and leading to arrangement drift. Even when the full vocal sequence is provided as conditioning, the model still makes myopic local decisions rather than globally coordinated refinements, which fundamentally limits its ability to maintain \textit{\textbf{Global Coherence}} and to plan the long-horizon structural development required for \textit{\textbf{Dynamic Orchestration}}.

Second, \textit{continuous-latent} (\textit{Cont. Lat.}) generation models like Diffusion Transformer (DiT)~\cite{peebles2023scalable}, exemplified by ACE-Step 1.5~\cite{gong2026ace} and SongEcho~\cite{li2026songecho}, alleviate the context limitations of AR systems by scaling more naturally to long-form generation. However, because they operate primarily in continuous latent spaces, they still face a major challenge in \textit{\textbf{Acoustic Authenticity}}. In practice, continuous latent modeling often struggles to faithfully reconstruct fine-grained acoustic details in dense accompaniment, leading to audible degradation in sound quality and, in some cases, a noticeable ``plastic'' texture. In addition, some methods rely on regenerated vocals to maintain structural consistency~\cite{li2026songecho}, which breaks the assumption of preserving the original raw vocal input. More broadly, this reveals a second challenge beyond generation quality: many existing V2A systems are not truly robust in zero-shot settings, as their performance depends heavily on auxiliary reference signals or intermediate reconstruction pipelines. Furthermore, while recent commercial \textit{Cont. Lat.} models like Suno V5  achieve high acoustic realism~\cite{suhailudheen2025suno}, their accompaniment generation workflows typically require manual adjustment  rather than functioning as fully end-to-end solutions.

Consequently, existing approaches face a clear trade-off: discrete autoregressive models better preserve local acoustic fidelity but are difficult to scale to song-level generation, while continuous-latent models improve long-range continuity at the cost of reduced acoustic authenticity. Therefore, an ideal V2A framework should not only resolve the trade-off among authenticity, coherence, and orchestration but also reduce dependence on auxiliary conditioning in order to remain effective in zero-shot accompaniment generation.

To address this trilemma, we introduce \textsc{LaDA-Band} (\textbf{La}nguage-based \textbf{D}iffusion \textbf{A}ccompaniment \textbf{Band}), an end-to-end framework that formulates full-song V2A generation as \textbf{Discrete Masked Diffusion}~\cite{chao2025beyond}, i.e., a global, non-autoregressive denoising formulation over discrete audio tokens. Rather than choosing between discrete autoregressive decoding and continuous-latent generation, \textsc{LaDA-Band} explores a new design point that combines the acoustic fidelity of discrete audio tokens with the full-sequence bidirectional iterative refinement capability of Language Diffusion Models~\cite{nielarge,zhu2025llada}.

This design addresses the Accompaniment Trilemma through a unified mechanism with clearly separated roles. 
First, LaDA-Band operates on discrete audio tokens rather than continuous latents, which preserves fine acoustic details through high-fidelity codec reconstruction and thus supports stronger \textit{\textbf{Acoustic Authenticity}}. 
Second, we formulate V2A generation as \textbf{Discrete Masked Diffusion}, a global, non-autoregressive denoising formulation. 
Unlike AR decoding, this formulation enables full-sequence bidirectional conditioning on the vocal track and iterative refinement over the entire accompaniment, thereby improving long-range structural consistency and synchronization, which strengthens \textit{\textbf{Global Coherence}}. In addition, to stabilize learning in accompaniment regions with weak or absent vocal anchoring (e.g., intros and interludes), we complement masked reconstruction with an auxiliary replaced-token discrimination objective that provides dense token-level supervision.
Third, to make \textbf{Discrete Masked Diffusion} effective for full-song accompaniment generation, we introduce a dual-track conditioning design and a two-stage progressive curriculum. Together, they extend the model from local vocal-accompaniment alignment to song-level arrangement evolution, thereby supporting stronger \textit{\textbf{Dynamic Orchestration}}.

In summary, our main contributions are threefold:
    
1. We formulate vocal-to-accompaniment generation as \textbf{Discrete Masked Diffusion}, a global, non-autoregressive denoising formulation that combines discrete audio token modeling with full-sequence bidirectional refinement.
    
2. To make this formulation effective for full-song V2A generation, we develop a dual-track conditioning architecture, together with an auxiliary dense token discrimination objective and a two-stage curriculum, to make Discrete Masked Diffusion effective for full-song V2A generation.

3. Experiments on both academic and real-world benchmarks show that \textsc{LaDA-Band} improves \textit{\textbf{Acoustic Authenticity}}, \textit{\textbf{Global Coherence}}, and \textit{\textbf{Dynamic Orchestration}} over existing baselines, while remaining robust under  zero-shot settings.

\begin{figure*}[t]
    \centering
    \includegraphics[width=0.98\linewidth]{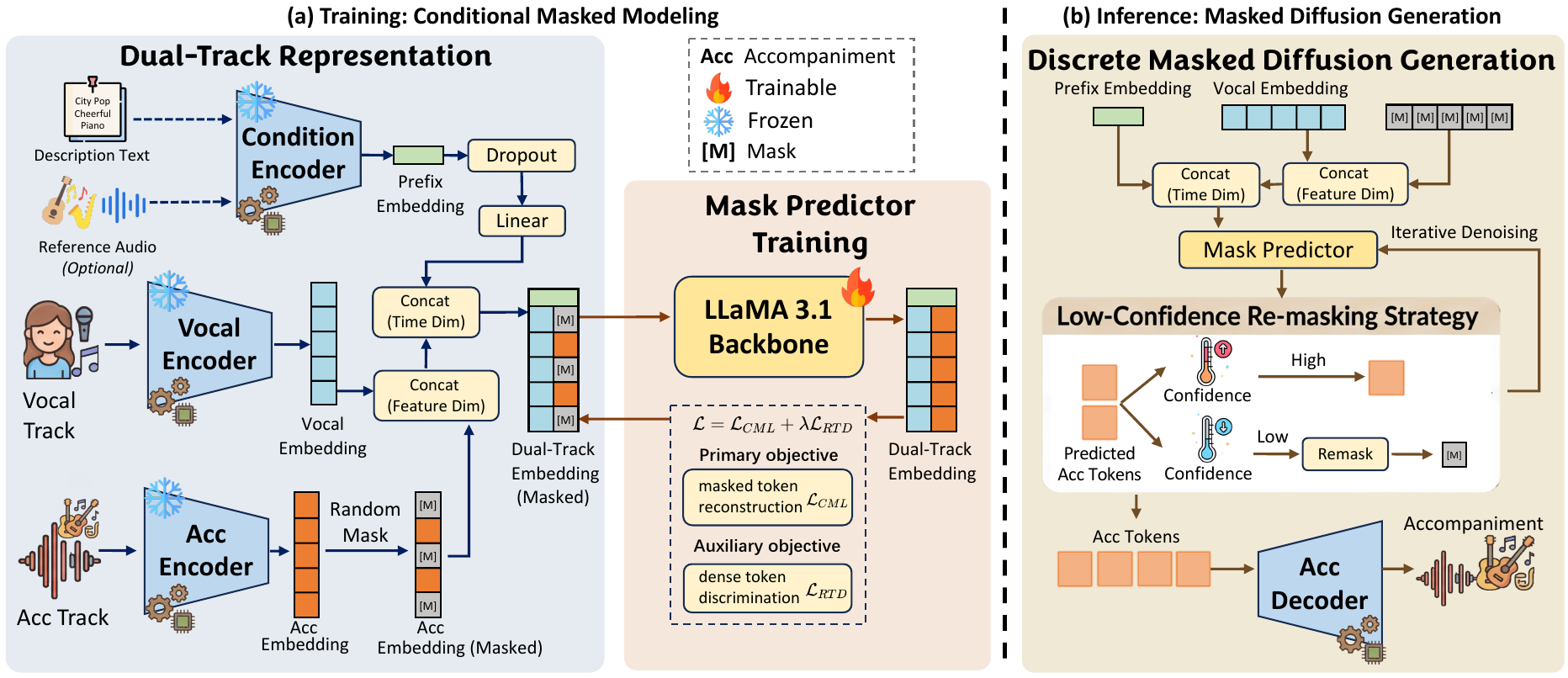}
\caption{\textbf{Overview of the \textsc{LaDA-Band} framework}, illustrating (a) the training of  LLaMA backbone  on  dual-track representations, and (b) the inference of \textbf{full-song} accompaniments via a  masked diffusion process. }
\label{fig:framework}
\end{figure*}

\section{Related Works}
\subsection{Text-to-Music Models}
\label{subsec:text_to_music}
Recent progress in text-to-music generation has substantially improved the fidelity, diversity, and temporal scale of controllable music synthesis~\cite{li2025survey}. Early systems were dominated by \textit{discrete autoregressive} generation, as exemplified by SongGen~\cite{liu2025songgen}, MusicGen~\cite{copet2023simple}, YuE~\cite{yuan2025yue}, LeVo~\cite{lei2025levo}, and HeartMuLa~\cite{yang2026heartmula}. More recent approaches have moved toward \textit{continuous-latent} generation using diffusion or flow-matching objectives, including JEN-1, Composer~\cite{liu2025jam}, DiffRhythm~\cite{jiang2025diffrhythm}, and hybrid systems such as SongBloom~\cite{yang2025songbloom} and ACE-Step~\cite{gong2025ace,gong2026ace}. In parallel, commercial foundation models such as Suno, Udio, Seed-Music~\cite{bai2024seed}, and Lyria 3 have pushed full-song generation to a new level of realism and scale.

Despite these advances, general-purpose text-to-music systems primarily target open-ended music synthesis from coarse semantic prompts, rather than tightly controlled accompaniment generation. As a result, they are not designed to align with arbitrary raw vocal tracks at fine temporal granularity, such as micro-level rhythmic synchronization, phrase-level harmonic support. In contrast, V2A generation requires the model to preserve the original vocal input while producing accompaniment that is  precisely synchronized.

\subsection{Vocal-to-Accompaniment Generation}
\label{subsec:accompaniment_gen}
Vocal-to-accompaniment (V2A) generation aims to synthesize instrumental accompaniment that strictly follows a given vocal input. One line of work extracts melody from vocals and then applies symbolic accompaniment generation models~\cite{wang2022songdriver,zhao2021accomontage,wang2023emotion,ou2024unifying,le2024meteor,choi2026d3pia} to produce MIDI drafts for subsequent production. While effective in controlled workflows, such pipelines require  manual intervention and thus have limited end-to-end practicality. Similarly, the Suno V5 requires manual adjustment to achieve the ideal accompaniment effect, especially for rock and DJ songs with complex mixing.

To move toward direct audio-level generation, early systems such as SingSong~\cite{donahue2023singsong}, FastSAG~\cite{chen2024fastsag} formulated the task as audio-to-audio translation, using source-separated datasets. Subsequent methods explored different combinations of representation space and generation formulation. SongEditor~\cite{yang2025songeditor} follows a \textit{discrete autoregressive} formulation based on neural audio codecs, while MuseControlLite~\cite{tsai2025musecontrollite}, AnyAccomp~\cite{zhang2025anyaccomp}, ACE-Step 1.5~\cite{gong2026ace}, and SongEcho~\cite{li2026songecho,hou2025editing} represent more recent \textit{continuous-latent} approaches based on diffusion or flow matching. Although these systems have improved controllability and generation quality, they still exhibit a common trade-off: discrete autoregressive methods preserve local acoustic detail but are difficult to scale to full-song generation, whereas continuous-latent methods scale more naturally to long-form generation but may sacrifice fine-grained acoustic fidelity.

Parallel efforts have also improved controllability and workflow integration. Melodist~\cite{hong-etal-2024-text} introduced tri-tower contrastive pretraining for text-guided accompaniment generation, while STAGE~\cite{strano2025stage} used prefix-based conditioning to generate specific stems synchronized with a click track. To further stabilize long-form generation, ComposerFlow~\cite{tsai2026midi} extracts explicit symbolic priors from vocals to guide acoustic synthesis. In contrast, our work targets end-to-end full-song V2A generation without relying on regenerated vocals, complex reference-audio pipelines, or  symbolic transcription.

\subsection{Language Diffusion Models}
Language Diffusion Models (LDMs) have recently emerged as a strong alternative to autoregressive modeling in NLP and multimodal generation. Early work such as MDLM~\cite{sahoo2024simple} and ADLM~\cite{routanchored} showed that discrete non-autoregressive diffusion models can approach, and in some settings surpass, autoregressive counterparts. More recently,  masked diffusion ~\cite{nielarge} language modeling can scale to LLaMA3-level performance at the 8B parameter regime. These ideas have also been extended to multimodal settings: LLaDA-V~\cite{you2025llada} incorporates visual conditioning for image-grounded dialogue, while DIFFA-style models~\cite{zhou2025diffa,tian2026dllm} use adapter-based designs to align speech and text for audio understanding. Other discrete masked diffusion variants have also been explored for language generation tasks such as sign-language translation and summarization~\cite{do2025discrete,demirag2024benchmarking}.

In the audio generation domain, masked modeling and non-autoregressive iterative decoding have been studied in systems such as SoundStorm~\cite{borsos2023soundstorm} and VampNet~\cite{garcia2023vampnet}. Although masked generation has shown promise in unconditional or text-conditioned audio synthesis, its potential for the improved version of diffusion and application in V2A generation remains underexplored, especially under the strict requirement of preserving alignment with a continuous, unmasked vocal track over full-song duration. Our work builds on the LDM formulation and introduces a discrete, global, non-autoregressive denoising formulation for zero-shot V2A generation based on dual-track representation.

\section{Task Definition}
\label{sec:task_definition}

The primary objective of our work is to generate an accompaniment sequence given a vocal track and a reference audio or descriptive text condition. We formulate this vocal-to-accompaniment generation as a conditional sequence modeling task. 


Let $\mathbf{V} = \{v_1, v_2, \dots, v_T\}$ and $\mathbf{A} = \{a_1, a_2, \dots, a_T\}$ denote the discrete vocal tokens and the corresponding accompaniment tokens, respectively. $T$ is the sequence length. Let $\mathbf{c}$ denote the global condition extracted from a reference audio or text. 

The goal of the conditional generative model is to learn the data distribution $P(\mathbf{A} \mid \mathbf{V}, \mathbf{c})$. Unlike traditional autoregressive (AR) models that factorize this joint probability strictly left-to-right as $\prod_{t=1}^T P(a_t \mid a_{<t}, \mathbf{V}, \mathbf{c})$, we adopt a Non-Autoregressive (NAR) masked generative formulation. We aim to model the joint distribution of masked accompaniment tokens directly, enabling highly parallel generation and global iterative refinement.

\section{Methodology}
\label{sec:method}

To address the conditional accompaniment generation task, we formulate \textsc{LaDA-Band} as a  Discrete Masked Diffusion framework for vocal-to-accompaniment generation. Concretely, our method combines an LLaMA-based Mask Predictor with a prefix-conditioned dual-track representation, enabling non-autoregressive generation with global iterative refinement.

\subsection{Prefix-based Conditional Dual-Track Representation}
\label{subsec:input_rep}

The input to our model consists of three components: the local vocal sequence $\mathbf{V}$, the \textbf{partially masked accompaniment sequence} $\tilde{\mathbf{A}}$, and the global condition $\mathbf{c}$. 

Inspired by multi-track tokenization~\cite{lin2025duo}, to sequence length $T$ while capturing both audio modalities, we utilize a dual-track concatenation strategy along the feature dimension. The discrete tokens of both tracks are mapped to continuous embeddings $\mathbf{E}_{voc} \in \mathbb{R}^{T \times \frac{D}{2}}$ and $\mathbf{E}_{acc} \in \mathbb{R}^{T \times \frac{D}{2}}$, respectively, where $D$ is the hidden dimension of the backbone model. The dual-track representation is then formulated as:
\begin{equation}
\mathbf{E}_{dual} = [\mathbf{E}_{voc} ; \mathbf{E}_{acc}] \in \mathbb{R}^{T \times D},    
\end{equation}
where $[ ; ]$ denotes concatenation along the feature dimension.

To inject the global text or audio condition $c$, we employ a compact prefix-conditioning strategy. Importantly, these prefix tokens are not intended to encode temporally localized arrangement plans across the full song. Instead, they provide low-bandwidth global guidance on style, instrumentation tendency, and reference bias, while orchestration is primarily modeled through the continuously visible vocal track, the dual-track sequence representation, and full-sequence bidirectional denoising. The CLaMP3~\cite{wu2025clamp}  condition vector is transformed via a linear projection layer into the backbone's hidden dimension, resulting in a prefix embedding $\mathbf{E}_{prefix} \in \mathbb{R}^{P \times D}$, where $P$ is the prefix length. This prefix is explicitly prepended to the dual-track sequence along the temporal dimension:
\begin{equation}
\mathbf{X}_{input} = [\mathbf{E}_{prefix} \parallel \mathbf{E}_{dual}] \in \mathbb{R}^{(P+T) \times D},    
\end{equation}
where $\parallel$ denotes concatenation along the time dimension. This structural design allows the Transformer backbone to attend to the global condition across all time steps via standard bidirectional self-attention mechanisms.

\subsection{Mask Predictor Training}
\label{subsec:mask_predictor}

The core of \textsc{LaDA-Band} is a Mask Predictor instantiated with LLaMA 3.1-1B~\cite{grattafiori2024llama}. Given the combined input sequence $\mathbf{X}_{input}$, it reconstructs the original accompaniment tokens from the masked accompaniment $\tilde{\mathbf{A}}$. Through bidirectional attention over the padded sequence of length $(P+T)$, the predictor captures fine-grained vocal--accompaniment alignment. The hidden states at accompaniment positions are projected to logits $\mathbf{L} \in \mathbb{R}^{T \times |\mathcal{V}_{acc}|}$, where $|\mathcal{V}_{acc}|$ is the accompaniment vocabulary size.

\textbf{Probabilistic Formulation and Forward Process.}
Given a ground-truth accompaniment sequence $\mathbf{A}_0$, we define a forward masking process that independently replaces each token $a_i$ with a special [MASK] token $M$ with probability $t \in (0,1]$, yielding a partially masked sequence $\tilde{\mathbf{A}}_t$. Following prior language diffusion modeling~\cite{you2025llada}, this masking process treats sequence information as approximately proportional to the number of observed tokens.

\textbf{Reverse Denoising and Conditional Masked Modeling.}
The reverse process is parameterized by a mask predictor $p_\theta(\cdot \mid \tilde{\mathbf{A}}_t, \mathbf{V}, \mathbf{c})$, which predicts all masked tokens simultaneously. We train it with a conditional masked modeling loss over the masked subset $\mathcal{M}_t$:
\begin{equation}
\mathcal{L}_{\mathrm{CML}}
=
-\mathbb{E}_{t,\mathbf{A}_0,\tilde{\mathbf{A}}_t}
\left[
\frac{1}{t \cdot T}
\sum_{i=1}^{T}
\mathbf{1}[\tilde{a}_t^{i}=M]
\log p_\theta(a_0^{i}\mid \tilde{\mathbf{A}}_t,\mathbf{V},\mathbf{c})
\right].
\end{equation}
This objective is a principled upper bound on the negative log-likelihood~\cite{you2025llada} and supports bidirectional,  coherent denoising.

\textbf{Auxiliary Replaced Token Detection.}
In practice, masked reconstruction alone is insufficient in accompaniment regions with weak or absent vocal anchoring, such as intros and interludes, where exact recovery is under-constrained and often yields overly sparse textures. To address this, we introduce an auxiliary Replaced Token Detection (RTD) objective~\cite{li2022pre}, which provides dense token-level supervision over the full sequence. Rather than forcing exact reconstruction everywhere, RTD encourages the model to judge whether each accompaniment token is contextually plausible. To avoid additional overhead, the Mask Predictor itself also serves as the discriminator.

Specifically, we replace accompaniment tokens with sampled alternatives from $\mathcal{V}_{acc}$. Let $d_i \in \{0,1\}$ indicate whether the token at position $i$ is original. The RTD objective is:
\begin{equation}
\mathcal{L}_{\mathrm{RTD}}
=
-\sum_{i=1}^{T}
\left[
d_i \log D_\theta(\tilde{\mathbf{A}}_t,i)
+
(1-d_i)\log\bigl(1-D_\theta(\tilde{\mathbf{A}}_t,i)\bigr)
\right],
\end{equation}
where $D_\theta(\cdot,i)$ denotes the probability that the $i$-th token is original. The final objective is
\begin{equation}
\mathcal{L}
=
\mathcal{L}_{\mathrm{CML}}
+
\lambda \mathcal{L}_{\mathrm{RTD}},
\end{equation}
where $\lambda$ balances reconstruction and discrimination. In summary, $\mathcal{L}_{\mathrm{CML}}$ learns masked token reconstruction, while $\mathcal{L}_{\mathrm{RTD}}$ regularizes full-sequence token plausibility, especially in weakly anchored accompaniment regions.

\subsection{Masked Diffusion-based Generation}
\textbf{Inference via Low-Confidence Remasking.} During sampling, we simulate the reverse trajectory from $t=1$ (fully masked) to $t=0$ (fully observed) under a cosine schedule~\cite{fox2026learning}. At each intermediate step from $t$ to $s < t$, the model predicts values for all currently masked positions. To maintain the integrity of the reverse Markov transition~\cite{zhu2025llada}, we adopt a low-confidence remasking strategy. Specifically, only the predicted tokens with the highest confidence scores $c_i = \max_{v \in \mathcal{V}_{acc}} p_\theta(a_i = v \mid \cdot)$ are retained, while the $s/t$ fraction of tokens with the lowest confidence are re-masked for the next iteration. This coarse-to-fine mechanism anchors structural milestones first and refines rhythmic nuances later, ensuring strict global alignment with the vocal track.

\subsection{Two-stage Curriculum training}

To effectively scale our model from local acoustic alignment to full-song generation with diverse stylistic capabilities, we streamline the training into a two-stage progressive curriculum, utilizing sequence length scaling and data selection to enhance final generation quality.

\textbf{Stage 1: Short-form Base Training.} In the initial phase, we train the model on the full large-scale dataset using short audio clips. This stage focuses on enabling the Mask Predictor to learn fundamental vocal-accompaniment temporal alignment, basic chord harmonization, and appropriate instrumentation. To prevent the model from overfitting to the global CLaMP3~\cite{wu2025clamp} condition and to encourage a deeper understanding of the vocal track's intrinsic structural cues, we introduce a random dropout strategy to the global style. This approach essentially acts as Classifier-Free Guidance~\cite{ho2022classifier}, ensuring robust accompaniment alignment even when textual prompts are weak or absent.

\textbf{Stage 2: Long-form High-Quality Training.} Generating coherent full songs requires handling highly diverse musical styles and complex arrangements. To ensure stable long-form generation and further enhance acoustic fidelity, we expand the training context to full-length audio and introduce optimized data quality filtering. Specifically, we implement a data selection pipeline to filter the training corpus. The filtering criteria encompass automated aesthetic evaluations via SongEval~\cite{yao2025songeval}, the detection and exclusion of acoustic artifacts introduced by vocal-accompaniment source separation models, data augmentation through speed and pitch modification, and the structural analysis of the accompaniment tracks. We also invited music professionals to conduct data screening based on musical structure and the listening experience of the intro. Detailed filtering procedures are provided in the appendix. The model is then fine-tuned for long-form generation exclusively on this refined subset. This combination of high-quality data and expanded context windows enables \textsc{LaDA-Band} to seamlessly adapt to multi-layered, full-song accompaniments while avoiding the generation of low-quality acoustic artifacts.

\section{Experiments}
\label{sec:experiments}

\begin{table*}[t]
\centering
\caption{Performance comparison of \textsc{LaDA-Band} against SOTA \textit{open-source} baselines under their representative full conditions on Suno70k (academic).  RTF denotes real-time factor ($\downarrow$). All inference was performed on a single 40GB A100 GPU}
\label{tab:main_results}
\setlength{\tabcolsep}{3.3pt}
\resizebox{\textwidth}{!}{
\begin{tabular}{ll ccc ccc ccc c}
\toprule
\multirow{2}{*}{\textbf{Method}} & \multirow{2}{*}{\textbf{Formulation}} & \multicolumn{3}{c}{\textbf{Acoustic Auth.}} & \multicolumn{3}{c}{\textbf{Global Coherence}} & \multicolumn{3}{c}{\textbf{Dynamic Orch.}} & \multirow{2}{*}{\textbf{RTF}$\downarrow$} \\
\cmidrule(lr){3-5} \cmidrule(lr){6-8} \cmidrule(lr){9-11}
& & FAD $\downarrow$ & Clar. $\uparrow$ & Nat. $\uparrow$ & Onset F1 $\uparrow$ & HPCP Sim. $\uparrow$ & Coh. $\uparrow$ & CLaMP $\uparrow$ & Mus. $\uparrow$ & Mem. $\uparrow$ & \\
\midrule
\multicolumn{12}{l}{\textit{Short-form (10s) Generation Baselines (Chunked / Truncated)}} \\
SongEdit-3.0B & Disc. AR & 31.40 & 3.42 & 3.35 & 41.2 & 86.4 & 2.85 & 4.10 & 3.12 & 3.05 & 1.42 \\
AnyAcc-3.7B & Cont. Lat. & 48.20 & 3.15 & 3.18 & 46.5 & 88.1 & 3.02 & 4.55 & 3.20 & 3.18 & 0.48 \\
MCLite-3.7B & Cont. Lat. & 52.10 & 3.10 & 3.12 & 51.3 & 89.5 & 3.15 & 4.85 & 3.25 & 3.22 & 0.39 \\
\midrule
\multicolumn{12}{l}{\textit{Long-form (> 4min) Full-Song Generation Baselines}} \\
SongEdit-3.0B & Disc. AR & - & - & - & - & - & - & - & - & - & OOM \\
ACE-Step-3.7B & Cont. Lat. & 46.50 & 3.23 & 3.25 & 60.4 & 93.2 & 3.44 & 5.20 & 3.27 & 3.38 & \textbf{0.07} \\
SongEcho-3.6B & Cont. Lat. & 45.90 & 3.54 & 3.49 & 67.0 & 94.5 & 3.85 & 10.20 & 3.57 & 3.69 & 0.31 \\
\midrule
\textbf{LaDA-Band-3.0B} & \textbf{Disc. Diff.} & \textbf{22.70} & \textbf{3.76} & \textbf{3.69} & \textbf{82.0} & \textbf{94.8} & \textbf{3.95} & \textbf{14.83} & \textbf{3.67} & \textbf{3.87} & \underline{0.26} \\
\bottomrule
\end{tabular} }
\end{table*}

\begin{table*}[t]
\centering
\caption{\textbf{Zero-shot robustness, and generalization from academic to real-world benchmarks.} We evaluate representative baselines and \textsc{LaDA-Band} under reduced-conditioning settings on Suno70k (academic) and IHP (real-world) to isolate robustness beyond full-strength system comparison. }
\label{tab:zeroshot_generalization}
\resizebox{\textwidth}{!}{
\begin{tabular}{ll cccc cccc}
\toprule
\multirow{2}{*}{\textbf{Method}} 
& \multirow{2}{*}{\textbf{Conditioning}} 
& \multicolumn{4}{c}{\textbf{Suno70k Academic}} 
& \multicolumn{4}{c}{\textbf{IHP Realworld}} \\
\cmidrule(lr){3-6} \cmidrule(lr){7-10}
& 
& CLaMP $\uparrow$
& HPCP Sim. $\uparrow$
& Onset F1 $\uparrow$
& FAD $\downarrow$
& CLaMP $\uparrow$
& HPCP Sim. $\uparrow$
& Onset F1 $\uparrow$
& FAD $\downarrow$ \\
\midrule

\multicolumn{10}{l}{\textit{Baselines}} \\
\midrule
ACE-Step-3.7B & Text + Ref.           & 5.20& 93.2 & 60.4 & 46.5 & 6.42 & 93.1 & 56.2 & 41.9 \\
ACE-Step-3.7B & Text (Zero-shot)            & 4.91 & 88.1& 55.4& 49.2& 6.61 & 92.5 & 55.3 & 41.9 \\
\hdashline
SongEcho-3.6B     & Text + Ref. + Lyrics  & 10.2 & 94.5 & 67.0 & 45.9 & 6.17 & 95.8 & 43.3 & 46.9 \\
SongEcho-3.6B     & Text + Lyrics    & 10.2& 90.2& 59.9& 56.9& 6.38 & 91.3 & 49.2 & 46.4 \\
SongEcho-3.6B     & Text  (Zero-shot) & 9.3 & 90.2 & 48.9 & 61.7 & 6.20 & 91.3 & 40.8 & 68.1 \\
\midrule

\multicolumn{10}{l}{\textit{Ours}} \\
\midrule
LaDA-Band-3.0B & Text + Ref.   & \textbf{14.8} & \textbf{94.8} & \textbf{82.0} & \textbf{22.7} & \textbf{7.65} & \textbf{96.3} & \textbf{68.2} & \textbf{31.7} \\
LaDA-Band-3.0B & Text (Zero-shot)  &  \textbf{13.1}& \textbf{91.8} & \textbf{60.5}& \textbf{22.7}& \textbf{7.10} & \textbf{94.8} & \textbf{67.1} & \textbf{32.5} \\
\bottomrule
\end{tabular}
}
\vspace{2pt}
\end{table*}

\subsection{Dataset and Preprocessing}
\label{subsec:dataset}

To support the two-stage curriculum training of \textsc{LaDA-Band}, we curated a massive-scale multi-modal music corpus. For the initial two stages—comprising foundation pre-training and paragraph-level alignment—we followed the data recipe of SongEditor \cite{yang2025songeditor} and the scale of ACE Step \cite{gong2026ace}, aggregating approximately 100k hours of diverse song data. In the data selection pipeline for Training Stage 2, we utilize the Suno70k  \cite{li2026songecho} and In-house Professional (IHP) dataset. The IHP dataset consists of recordings from a mainstream music platform (including user-recorded dry vocals), ensuring the model's adaptability to real-world, professionally produced content. Specific information and download channels for the IHP dataset can be found in the appendix. From both Suno70k and IHP, we reserved 1,000 samples as held-out test sets to evaluate the model's generalization across different data distributions, while the remainder were used for style-specific adaptation. This contains a total of around 140k samples with a total duration of around 10k hours. Similar to Suno70k, the IHP dataset is a text description with instrumentation, emotion, and style, annotated by Qwen2.5-Audio~\cite{chu2024qwen2}.

The preprocessing pipeline is designed to transform raw audio into high-quality discrete representations. We first employ BSRNN \cite{luo2023music}, a SOTA music source separation model, to decouple the vocal tracks from their corresponding accompaniments. The isolated vocals serve as the conditional guidance, while the accompaniments are designated as the generation targets. To bridge the gap between continuous waveforms and discrete generative modeling, all audio signals are tokenized using MuCodec \cite{xu2025mucodec}.

\subsection{Implementation Details}

The backbone is  instantiated from the LLaMA-3.2-1B configuration. The hidden dimension of this backbone is 2048. Text and reference-audio embeddings extracted by a frozen CLaMP3 encoder are linearly projected from 768 dimensions to the backbone latent space. The prefix length is fixed to $P=2$, corresponding to one text prefix token and one reference-audio prefix token. We  keep this prefix compact because these conditions serve as weak global controls rather than temporally aligned supervision. In our setting, song-level arrangement is primarily learned from the vocal-conditioned dual-track sequence instead of being explicitly encoded in the condition prefix. The text only contains styles, instrumental, and emotional descriptions.

All audio signals are resampled to 48~kHz. During training, we randomly crop a 40.96s (stage 1) / 163.84s (stage 2)  vocal--accompaniment segment pair and independently sample a 20.48~s accompaniment reference segment from the same song. To prevent data leakage, the reference audio and accompaniment fragments cut from each song have no overlap.
The reference audio is converted to mono before being encoded by CLaMP3. We apply frequency-balance augmentation (See appendix for details) to the accompaniment and reference streams with probability 0.5, while leaving the vocal stream unchanged. The MuCodec front-end encodes stereo audio into a single stream of discrete tokens at 25~Hz. The MuCodec codebook size is $1\times16384$, yielding a bitrate of $\approx$0.35~kbps. For accompaniment modeling, we reserve one mask token and one padding token, i.e., $acc\_mask=16384$ and $acc\_pad=16385$.

The CLaMP3 encoder is kept frozen throughout training. Raw reference audio is  mapped to a global 768-dimensional CLaMP3 embedding. During training, classifier-free conditioning dropout is applied independently to the text and audio prefixes with probability 0.5. MuCodec is also frozen. 

For discrete mask diffusion training, we sample a quasi-random progress variable $r \in [0,1]$ from a Sobol engine and convert it into a mask ratio using a cosine schedule. The RTD loss weight  $\lambda $ is set to 0.2, with argmax token replacement, and temperature 1.0.

Optimization uses AdamW with $\epsilon=10^{-8}$, and weight decay of 0.01. The learning rate is set to $1\times10^{-5}$ with a cos schedule and 10000 warmup steps. Training uses bf16-mixed precision, gradient clipping of 1.0, a per-device batch size of 2, and gradient accumulation over 4 steps. All training processes were completed on 8 $\times$ A800 (80G) GPUs. Stage 1 was trained for 150k steps, and Stage 2 was trained for 40k steps.

During inference, we use a cosine remasking schedule 60 denoising iterations with top-k=100, top-p=0.9, temperature 1.0, and mask temperature 10.5. The generated discrete accompaniment codes are then converted back to a waveform by a frozen MuCodec flow-matching decoder~\cite{xu2025mucodec}.

\subsection{Evaluation Metrics}
To comprehensively assess the generated accompaniments and validate our resolution of the Accompaniment Trilemma, we employ a combination of objective acoustic/alignment metrics and automated aesthetic evaluations. RTF values~\cite{chen2025f5} are intended to provide an inference efficiency comparison. (SongEval~\cite{yao2025songeval}).

To evaluate \textit{\textbf{Acoustic Authenticity}}, we utilize FAD~\cite{gui2024adapting} with clap embeddings ~\cite{elizalde2023clap} that measures the distribution gap between generated and real audio embeddings. This measurement is further complemented by the \textit{Clarity} and \textit{Naturalness} aesthetic metrics from SongEval.
For \textit{\textbf{Global Coherence}}, strict micro-temporal alignment with the vocal track and the execution of long-form structural progression without temporal drift are critical. We measure micro-level temporal synchronization using Onset F1~\cite{mcfee2015librosa}, and evaluate overall harmonic stability across the track using HPCP Sim. Additionally, the automated SongEval \textit{Coherence} metric is utilized to assess macroscopic structural integrity.
To assess \textit{\textbf{Dynamic Orchestration}}, we leverage the CLaMP metric, which evaluates how accurately the generated  audio aligns with the  styles, instrumental and emotional descriptions. This is further evaluated using SongEval's \textit{Musicality} and \textit{Memorability} metrics.

\subsection{Baselines}
We compare \textsc{LaDA-Band} against a representative set of state-of-the-art \textit{open-source} accompaniment generation baselines. Detailed reproduction settings are deferred to the appendix. Baselines are reproduced with matched training data. All reference audio uses isolated accompaniment; for methods requiring lyrics input, missing annotations are obtained with Qwen3-ASR~\cite{shi2026qwen3}.

\textbf{Short-form baselines.}
We evaluate SongEditor~\cite{yang2025songeditor} as the representative discrete AR baseline, denoted as SongEdit-3.0B, since it is reproduced at the same overall scale as LaDA-Band-3.0B. We further include two representative continuous-latent short-form baselines, AnyAccomp~\cite{zhang2025anyaccomp} and MuseControlLite~\cite{tsai2025musecontrollite}, denoted as AnyAcc-3.7B and MCLite-3.7B, respectively. Under our unified setup, MuseControlLite is reproduced on the ACE-Step backbone following the SongEcho protocol, while AnyAccomp retains its quantized melodic bottleneck design at a comparable overall scale. As these methods are limited by short context windows, they generate accompaniment, often leading to weaker song-level consistency. For these baselines, all evaluation samples are taken from the first 10 seconds of the first verse detected by SongFormer~\cite{hao2025songformer}.

\textbf{Long-form full-song baselines.}
We further compare with continuous-latent models designed for longer-context or full-song accompaniment synthesis, including ACE-Step 1.5~\cite{gong2025ace} and SongEcho~\cite{li2026songecho}, denoted as ACE-Step-3.7B and SongEcho-3.6B under our unified counting protocol. ACE-Step 1.5 serves as a strong latent baseline with  long-context generation and editing capability, while SongEcho represents a melody-conditioned full-song framework built on ACE-Step. We also evaluate the instrumental generation function of Suno V5 in the appendix, and report only its fully automatic end-to-end results.

\begin{figure}[t]
  \centering
  \includegraphics[width=\linewidth]{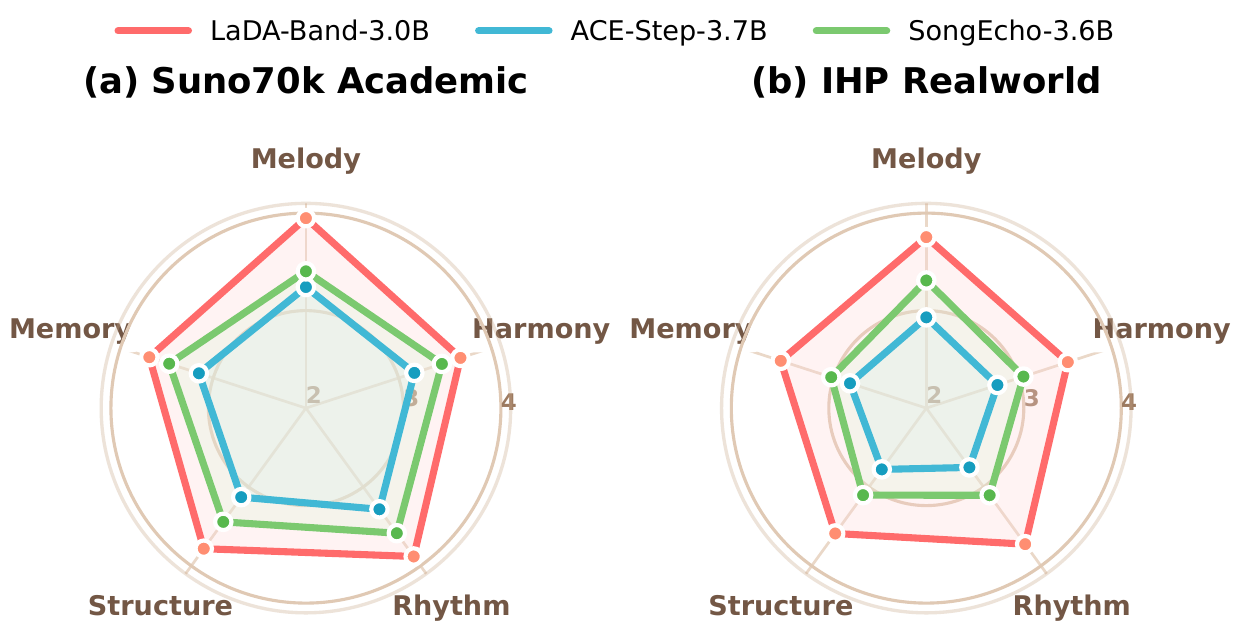}
  \caption{Comparison of LaDA-Band and continuous-latent baselines on the five SongEval dimensions \textbf{under the  restricted Zero-shot (Text-Only) conditioning scenario}.}
  \label{fig:songeval}
\end{figure}

\subsection{Main Objective Results}

\begin{table}[t]
\centering
\caption{\textbf{Ablation study of the proposed formulation and curriculum.} We replace core components and training stages to validate the gains of \textsc{LaDA-Band}.}
\label{tab:ablation_necessity}
\small
\setlength{\tabcolsep}{4pt}
\begin{tabular}{lccc}
\toprule
\textbf{Setting} & \textbf{Onset F1} & \textbf{FAD} & \textbf{CLaMP} \\   
\midrule
\textbf{Full LaDA-Band} & \textbf{82.0} & \textbf{22.70} & \textbf{14.83} \\
\ \ w/o Stage 2 (Stage 1 only) & 78.4 & 25.87 & 11.86 \\
\ \ w/o Discrete (Cont. latents) & 71.4 & 38.65 & 10.20 \\
\ \ w/o NAR (AR decoding) & 14.2 & 26.1 & 2.12 \\
\ \ w/o Dual-track (Single-stream) & 28.7 & 29.35 & 4.55 \\
\ \ w/o RTD Loss & 76.5 & 25.21 & 16.71 \\
\bottomrule
\end{tabular}
\end{table}

\begin{acks}
To Robert, for the bagels and explaining CMYK and color spaces.
\end{acks}


\begin{figure}[t]
  \centering
  \includegraphics[width=0.9\linewidth]{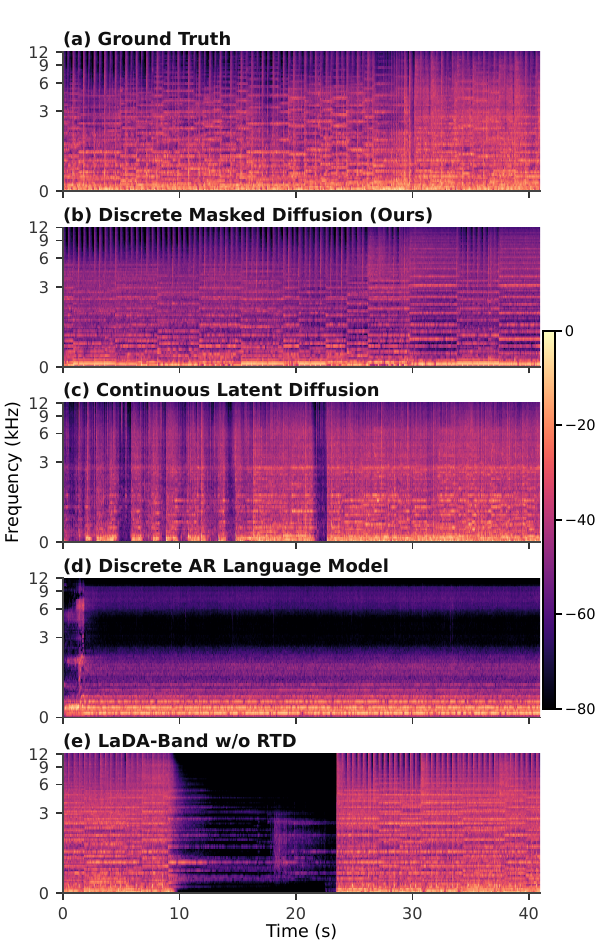}
  \caption{Spectrogram comparison of modeling formulations for long-form  accompaniment  generation}
  \label{fig:formulation_comparison}
\end{figure}
\textbf{LaDA-Band outperforms all competing baselines across all metrics corresponding to the Accompaniment Trilemma.}
It breaks the inherent trade-off between discrete autoregressive and continuous-latent models, surpassing state-of-the-art baselines on all core metrics of acoustic authenticity, global coherence, and dynamic orchestration in the full-song generation evaluation on the Suno70k benchmark. Furthermore, we ablated other components built upon our formulation (Section \ref{subsec:ablation}), and the performance still outperformed the baselines.

\textbf{LaDA-Band achieves a superior trade-off between inference efficiency and generation performance.}
With a real-time factor (RTF) of 0.12, it delivers comprehensive performance gains with lower latency than the prior state-of-the-art SongEcho-3.6B, while avoiding the out-of-memory (OOM) failure of the autoregressive SongEdit-3.0B in full-song generation.

\subsection{Zero-shot Robustness and Real-World Generalization}
\textbf{LaDA-Band maintains robust generation performance under the zero-shot text-only setting.}
Unlike baselines that suffer severe performance degradation when removing reference audio, it preserves core temporal synchronization and acoustic fidelity via conditioning dropout and dual-track modeling, outperforming  baselines under the  zero-shot setting on the Suno70k.

\textbf{LaDA-Band’s zero-shot performance even surpasses representative baselines under their full conditioning settings.}
On the real-world IHP dataset, its core metrics including FAD, Onset F1 and HPCP similarity under the zero-shot text-only setting comprehensively outperform those of ACE-Step and SongEcho with complete multi-modal conditioning, demonstrating strong real-world generalization. 

\subsection{Ablation Study}
\label{subsec:ablation}

Table~\ref{tab:ablation_necessity} validates how each core component of \textsc{LaDA-Band} resolves the Accompaniment Trilemma on the Suno70k benchmark.

\textbf{Discrete Tokens for \textit{Acoustic Authenticity}.}
Replacing discrete codec tokens with continuous latents (via a DiT backbone and Music-DCAE~\cite{gong2025ace}) causes blurring of high and low frequencies (see the zoomed-in version in Figure ~\ref{fig:teaser}). Although Music-DCAE can ensure high-fidelity reconstruction ability on text-to-music tasks, it shows that, in our V2A setting, discrete codec tokens are better suited to preserving acoustic fidelity under the proposed formulation due to the task's trilemma.

\textbf{Masked Diffusion for \textit{Global Coherence}.} Switching from NAR to AR decoding collapses the Onset F1 score. This empirical gap proves AR models suffer from compounding errors and arrangement drift in  contexts. 

\textbf{Dual-track $\&$ Curriculum for \textit{Dynamic Orchestration}.} Treating vocals as a single-stream prefix~\cite{yang2025songeditor} rather than our dual-track concatenation undermines temporal synchronization. Furthermore, removing the Stage 2 long-form training drops the CLaMP score. 

\textbf{If RTD Loss is not used, the intro and interlude are relatively empty}, and the instrumental texture is simple, which in turn affects the performance. For intro/interlude regions, where vocal guidance is weak and reconstruction-only, training tends to produce overly sparse accompaniment. 

\textbf{Even without RTD Loss and Stage 2 training, Discrete Masked Diffusion still maintains its advantages}. It remains superior to the Cont. Lat. baselines in terms of objective metrics. This proves the inherent superiority of our formulation.

Visualized in Fig.~\ref{fig:formulation_comparison}, \textsc{LaDA-Band} (b) faithfully reconstructs the Ground Truth (a) in complex harmonic regions. The continuous baseline (c) exhibits visible  blurring, losing instrumental crispness. The AR baseline (d) shows severe structural drift, reflecting an inability to plan music structures. \textsc{LaDA-Band} better preserves spectral detail to ensure acoustic authenticity.  (e) shows that without RTD loss, the intro/interlude sections lack supervision. As the vocal signal is sparse, the generated accompaniment regions also become sparse accordingly.

\subsection{Subjective Evaluation}
\label{subsec:subjective}
\label{sec:results}

\begin{table}[t]
\centering
\caption{Subjective MOS Results on Various Style Samples. All models adopt representative full conditioning. The 95\% confidence interval for the mean.}

\label{tab:mos}
\resizebox{0.48\textwidth}{!}{
\begin{tabular}{llccc}
\toprule
\textbf{Method} &  \textbf{AQ} & \textbf{MA} & \textbf{ON} & \textbf{Overall} \\
\midrule
ACE-Step-3.7B &  $2.217_{\pm 0.185}$ & $2.283_{\pm 0.198}$ & $2.017_{\pm 0.175}$  & $2.172_{\pm  0.155}$\\
SongEcho-3.6B &  $2.908_{\pm 0.172}$ & $3.354_{\pm 0.187}$ & $3.400_{\pm 0.169}$ & $	3.221_{\pm  0.153}$ \\
\midrule
\textbf{LaDA-Band-3.0B}  & \textbf{3.338}$_{\pm 0.184}$ & 	\textbf{3.367}$_{\pm 0.187}$ & \textbf{3.483}$_{\pm 0.179}$ & \textbf{3.396}$_{\pm  0.154}$\\
\bottomrule
\end{tabular}
}
\end{table}
The double-blind Mean Opinion Score (MOS) test evaluated 20 samples from the Suno70k test set (mainstream music styles). 17 participants (7 professionals, 10 amateurs) rated acoustic quality (AQ), musical alignment (MA), and orchestration naturalness (ON) on a 5-point scale.  
Our questionnaire design for the subjective evaluation process can be found in the appendix. \textsc{LaDA-Band} outperforms baselines across all, particularly in AQ. These results indicate advantages in perceived timbral quality and arrangement naturalness. Regarding AQ and MA, LaDA-Band and  SongEcho both have failure cases, and their metrics are similar.

\textbf{We observed that the main failure cases occurred in the jazz and funk genres}, because these types of music do not have coherence as a strict artistic standard. In genres like rock, ballad, and pop, which demand high coherence, LaDA-Band performs better due to its superior audio fidelity and orchestration.
To better contextualize  bad cases, we further analyze genre-specific behavior and representative failure cases in the appendix.

\subsection{Model Scaling and Sampling Strategy}
We scaled up the Mask Predictor to 3B parameters, which further improved performance. Experiments also verified that a cosine schedule with 20 denoising steps achieves the best efficiency-performance trade-off. Results are shown in the appendix.

\section{Conclusion}
We present \textsc{LaDA-Band}, an end-to-end framework that formulates full-song V2A generation as Discrete Masked Diffusion. This formulation, together with dual-track representation, auxiliary RTD loss, and training curriculum, provides a systematic solution to the Accompaniment Trilemma. Experiments on both academic and real-world benchmarks show consistent gains in acoustic authenticity, global coherence, and dynamic orchestration, while preserving robustness under zero-shot settings.

\section{Limitations and Future Work}
\textsc{LaDA-Band} still has several limitations. First, its performance depends on the source separation and audio codec pipeline, so upstream errors may propagate to the generated accompaniment. Second, it still offers limited fine-grained control over arrangement details such as instrument-level editing. Third, some stylistically free-form genres remain challenging. In the future, we will add richer conditioning, such as cross-attention to editing instructions, which could further improve fine-grained  controllability. This is complementary to our current goal of zero-shot end-to-end V2A generation under weak global conditions.

\bibliographystyle{ACM-Reference-Format}
\bibliography{sample-base}
\newpage
\appendix

\section{Training Corpus and Data Recipe}
\subsection{Introduction to Training Corpus}
All methods are reproduced on the same unified training corpus. 
Our training recipe follows the large-scale full-song setting of ACE-Step and further incorporates the \textsc{Suno70k} dataset introduced by SongEcho. 
We additionally build an in-house benchmark from large-scale commercial music and social singing platforms, covering both studio-quality songs and user-recorded singing content. 
This benchmark is notably more challenging than academic datasets, as it contains larger variation in vocal recording conditions, accompaniment styles, mixing quality, genre distribution, and song structure. 
As a result, it better reflects practical zero-shot V2A scenarios beyond relatively clean academic settings. 
The IHP dataset can be obtained by contacting the authors or requested on the demo website we maintain upon publication.

\subsection{Data Filtering and Augmentation Process}

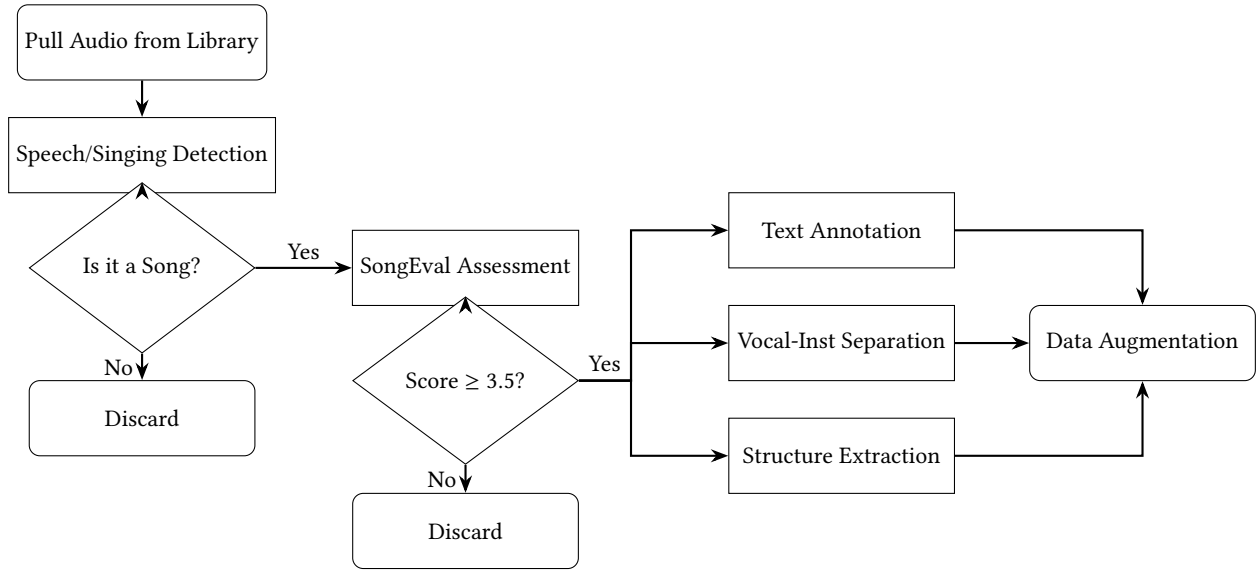
\begin{figure*}[h!]
\centering
\caption{Data Filtering and Augmentation Flowchart}
\label{fig:audio_pipeline}
\begin{tikzpicture}[
    startstop/.style={rectangle, rounded corners, minimum width=3cm, minimum height=1cm, text centered, draw=black, fill=white},
    process/.style={rectangle, minimum width=3cm, minimum height=1cm, text centered, draw=black, fill=white},
    decision/.style={diamond, minimum width=3cm, minimum height=1cm, text centered, draw=black, fill=white},
    arrow/.style={thick,->,>=Stealth},
    node distance=1.5cm and 1.2cm
]

\node (start) [startstop] {Pull Audio from Library};
\node (detection) [process, below of=start] {Speech/Singing Detection};
\node (is_song) [decision, below of=detection] {Is it a Song?};
\node (discard1) [startstop, below of=is_song, yshift=-0.5cm] {Discard};
\node (songeval) [process, right of=is_song, xshift=2.8cm] {SongEval Assessment};
\node (quality_check) [decision, below of=songeval] {Score $\geq 3.5$?};
\node (discard2) [startstop, below of=quality_check, yshift=-0.5cm] {Discard};

\node (text_ann) [process, right of=quality_check, xshift=3.5cm, yshift=2cm] {Text Annotation};

\node (separation) [process, below of=text_ann] {Vocal-Inst Separation};
\node (structure) [process, below of=separation] {Structure Extraction};
\node (aug) [startstop, right of=separation, xshift=2.5cm] {Data Augmentation};

\draw [arrow] (start) -- (detection);
\draw [arrow] (detection) -- (is_song);
\draw [arrow] (is_song) -- node[anchor=east] {No} (discard1);
\draw [arrow] (is_song) -- node[anchor=south] {Yes} (songeval);
\draw [arrow] (songeval) -- (quality_check);
\draw [arrow] (quality_check) -- node[anchor=east] {No} (discard2);
\draw [arrow] (quality_check) -- node[anchor=south] {Yes} ++(2.2cm,0) |- (text_ann);

\draw [arrow] (quality_check) -- ++(2.2cm,0) |- (separation);
\draw [arrow] (quality_check) -- ++(2.2cm,0) |- (structure);
\draw [arrow] (text_ann) -| (aug);
\draw [arrow] (separation) -- (aug);
\draw [arrow] (structure) -| (aug);

\end{tikzpicture}
\end{figure*}
 
The pipeline implements end-to-end quality control and feature extraction. The workflow follows a gate-controlled linear structure with parallelized feature extraction for efficiency, consisting of five core sequential stages. The pipeline initiates by pulling unprocessed audio files from the internal music library as the input source, with no preprocessing applied at this stage. Next, a binary classification model is deployed to distinguish between singing (song content) and speech (e.g, talk shows, podcasts); if the audio is classified as non-song (speech-dominant), it is immediately discarded, and only confirmed song content proceeds to the next stage. The `SongEval` system then evaluates audio quality across multiple sub-metrics such as clarity, noise level, and musical completeness; if the overall sub-item score is below 3.5, the audio is discarded as low-quality, and only songs with a score greater than or equal to 3.5 proceed to feature extraction. Qualified songs undergo parallel multi-dimensional feature extraction to generate structured metadata, including text annotation (transcription and semantic labeling of lyrics/vocal content) by Qwen-Audio 2.5, vocal-instrumental separation (isolation of vocal track from instrumental accompaniment) by BSRNN, and structure extraction (analysis and segmentation of musical structure such as verse, chorus, and bridge) by SongFormer. Finally, 20\% of the annotated dataset is augmented via $\times1.1 / \times 1.2 \times0.8 / \times 0.9$ time stretching and $\pm1 / \pm2$ pitch shifting to expand the dataset size and improve model generalization. If the intro/interlude is too long, the structure lacks a chorus, or podcast content is mixed in, such data will be manually filtered out based on the structural annotation information.

During training, we apply a stochastic frequency-balance augmentation to the accompaniment and reference audio, while leaving the vocal signal unchanged. For each sample, the augmentation is activated with probability $p=0.7$. When activated, we sequentially apply three biquad filters with independently sampled parameters: (i) a bass tone-control filter with gain $g_{\mathrm{bass}} \sim \mathcal{U}(-5,-2)\,\mathrm{dB}$, center frequency $f_{\mathrm{bass}} \sim \mathcal{U}(80,180)\,\mathrm{Hz}$, and quality factor $Q_{\mathrm{bass}} \sim \mathcal{U}(0.6,1.1)$; (ii) a mid-band peaking equalizer with gain $g_{\mathrm{mid}} \sim \mathcal{U}(1,3)\,\mathrm{dB}$, center frequency $f_{\mathrm{mid}} \sim \mathcal{U}(1200,2600)\,\mathrm{Hz}$, and quality factor $Q_{\mathrm{mid}} \sim \mathcal{U}(0.6,1.2)$; and (iii) a treble tone-control filter with gain $g_{\mathrm{treble}} \sim \mathcal{U}(1,3.5)\,\mathrm{dB}$, center frequency $f_{\mathrm{treble}} \sim \mathcal{U}(3200,5800)\,\mathrm{Hz}$, and quality factor $Q_{\mathrm{treble}} \sim \mathcal{U}(0.6,1.0)$. This augmentation simulates realistic spectral coloration by slightly attenuating low frequencies while emphasizing mid and high frequencies, thereby improving robustness to timbral and equalization variations in accompaniment-related inputs. After filtering, the audio is loudness-normalized to a target RMS level of $-16\,\mathrm{dB}$, with the peak level capped at $-1\,\mathrm{dB}$, so that the augmentation primarily changes spectral balance rather than overall loudness.

\section{Experimental Details}
\subsection{Subjective Blind Listening Test Design}
To quantitatively evaluate the subjective performance of the generated audio and guarantee the credibility of experimental results, this study designed a double-blind randomized subjective listening test, which adopts a 5-point Mean Opinion Score (MOS) scale allowing decimal scores based on raters' authentic listening perception. The unified rating criteria are defined as follows: 5 points for Excellent, 4 points for Good, 3 points for Fair, 2 points for Poor, and 1 point for Bad. Each audio sample is independently scored from three core evaluation dimensions, with detailed specifications as below. The first dimension is audio quality, which focuses on the presence of unnatural "plastic" texture, clarity of high-frequency components, impact of low-frequency components, and overall audio purity; a 5-point performance corresponds to natural instrumental timbre with clear and clean sound, while a 1-point performance corresponds to muffled high frequency, severe clipping and obvious distortion. The second dimension is harmony and alignment, which focuses on vocal intonation accuracy, harmony consistency between vocal and accompaniment, rhythmic alignment, and the presence of temporal drift; a 5-point performance corresponds to professional-level vocal output with harmonious listening experience and perfectly aligned rhythm, while a 1-point performance corresponds to disharmonious mix, chaotic rhythm, out-of-tune vocals or incorrect pitch. The third dimension is arrangement, which focuses on the natural evolution of accompaniment instrumentation, musical expressiveness, and avoidance of mechanical repetition; a 5-point performance corresponds to rich instrumentation, clear hierarchical structure of musical segments, and accurate matching with vocal emotion and musical style, while a 1-point performance corresponds to monotonous instrumentation, lack of musical hierarchy, and no clear style orientation.

For the test procedure, the playback order of all audio samples is fully randomized for each individual participant, which effectively eliminates potential order effects and sequence bias on rating results. This test strictly follows the double-blind principle to maximize the objectivity and reliability of scoring data. Specifically, all participants are completely unaware of the correspondence between audio samples and generation models throughout the test. All audio files are renamed with randomly generated unique character strings, and only the test administrator has access to the mapping between anonymous filenames and corresponding models, which completely avoids interference from raters' subjective preference for specific models on scoring results.

For participant composition, all recruited raters have normal bilateral hearing ability and basic music listening experience, to ensure they can accurately perceive subtle differences across the three evaluation dimensions. Detailed demographic information of participants, including age, gender, music professional background and daily audio listening habits, will be collected and statistically summarized in the complete experimental analysis. Rating data is collected through a standardized questionnaire template, and the full scoring record sheet is designed as shown in Table \ref{tab:blind_listening_score_sheet}, which contains the anonymous filename of the audio sample, unique song ID, and dedicated scoring columns for the three evaluation dimensions.

\begin{table*}[htbp]

  \centering
  \caption{Audio File Metadata for Double-Blind Supervisor Review}
  \label{tab:audio-files}
  \begin{tabular}{l l l l l}  
    \toprule
    Filename & Song ID & Audio Quality & Harmony \& Alignment & Arrangement \\
    \midrule
    0fc13f0f-jhl6ca.mp3 & 0fc13f0f &  &  &  \\
    0fc13f0f-plvvpo.mp3 & 0fc13f0f &  &  &  \\
    0fc13f0f-wn5w6r.mp3 & 0fc13f0f &  &  &  \\
    \bottomrule
  \end{tabular}
\end{table*}
\subsection{Baseline Reproduction Details}
We compare against representative open-source baselines spanning both discrete autoregressive and continuous-latent paradigms, including SongEditor (AAAI 2025), SongEcho (ICLR 2026), ACE-Step 1.5, and MuseControlLite (ICML 2025). 
These baselines cover substantially different generation formulations, representation spaces, and conditioning mechanisms. The components are detailed in Table ~\ref{tab:baseline_arch_summary}. When counting the parameters of each model, all encoders, backbones, and decoders are included.

For fair comparison, all baselines are reproduced on our unified training corpus with matched training data whenever possible. 
While different paradigms require model-specific training recipes, we standardize the evaluation pipeline and the core conditioning inputs across methods, including extracted vocals and textual prompts, to reduce data- and protocol-induced bias. 
All reference-audio conditions use isolated accompaniment. 
For methods requiring lyrics input, missing annotations are obtained using Qwen3-ASR. 
We also adopt a unified total-parameter counting protocol that includes the backbone and all auxiliary modules.

SongEditor is reproduced by following its original discrete-token autoregressive formulation and adapting it to our V2A setting with autoregressive masking and decoding. 
ACE-Step 1.5 is reproduced following its original continuous-latent recipe based on Music-DCAE and a linear DiT backbone. 
SongEcho is reproduced on top of the ACE-Step backbone while retaining its original IA-EiLM melody-conditioning design. 
MuseControlLite is reproduced under the same ACE-Step-based protocol as SongEcho for a fair comparison among controllable continuous-latent baselines.

\begin{table*}[t]
\centering
\caption{\textbf{Architectural summary of reproduced baselines.} 
For SongEcho and MuseControlLite, we report the architecture under our ACE-Step-based reproduction protocol.}
\label{tab:baseline_arch_summary}
\setlength{\tabcolsep}{5pt}
\renewcommand{\arraystretch}{1.15}
\begin{tabular}{lccc}
\toprule
\textbf{Method} & \textbf{Encoder} & \textbf{Backbone} & \textbf{Decoder} \\
\midrule

SongEditor
& \begin{tabular}[c]{@{}l@{}}MuCodec encoder\end{tabular}
& \begin{tabular}[c]{@{}l@{}}Autoregressive\\ LLaMA 3.1-1B\end{tabular}
& \begin{tabular}[c]{@{}l@{}}MuCodec decoder\end{tabular} \\

\hdashline

ACE-Step 1.5
& \begin{tabular}[c]{@{}l@{}}Music-DCAE encoder\\ frozen mT5-base text encoder\\ SongGen-style lyric encoder\\ PLR-OSNet-style speaker encoder\end{tabular}
& \begin{tabular}[c]{@{}l@{}}Linear DiT\end{tabular}
& \begin{tabular}[c]{@{}l@{}}Music-DCAE decoder\\ universal music vocoder\end{tabular} \\

\hdashline

SongEcho
& \begin{tabular}[c]{@{}l@{}}Music-DCAE encoder\\ frozen mT5-base text encoder\\ SongGen-style lyric encoder\\ PLR-OSNet-style speaker encoder\\ RVMPE pitch extractor\\ melody encoder\end{tabular}
& \begin{tabular}[c]{@{}l@{}}ACE-Step Linear DiT\\ with IA-EiLM\end{tabular}
& \begin{tabular}[c]{@{}l@{}}Music-DCAE decoder\\ universal music vocoder\end{tabular} \\

\hdashline

MuseControlLite
& \begin{tabular}[c]{@{}l@{}}Music-DCAE encoder\\ frozen mT5-base text encoder\\ SongGen-style lyric encoder\\ PLR-OSNet-style speaker encoder\\ lightweight control encoder\end{tabular}
& \begin{tabular}[c]{@{}l@{}}ACE-Step Linear DiT\\ with decoupled cross-attention adapters\end{tabular}
& \begin{tabular}[c]{@{}l@{}}Music-DCAE decoder\\ universal music vocoder\end{tabular} \\

\bottomrule
\end{tabular}
\end{table*}

\subsection{Real-Time Factor (RTF)}
We report Real-Time Factor (RTF) as
\begin{equation}
\mathrm{RTF} = \frac{T_{\mathrm{infer}}}{T_{\mathrm{audio}}},
\end{equation}
where $T_{\mathrm{infer}}$ denotes the end-to-end wall-clock inference time, and $T_{\mathrm{audio}}$ denotes the duration of the generated accompaniment audio. 

For fair comparison, all reported RTF values are measured in an end-to-end manner, including both input representation encoding and waveform decoding stages, rather than counting only the backbone generation time. Therefore, the reported RTF reflects the practical deployment efficiency of the full system.

\section{Supplementary Experiment}
\subsection{Why Suno Is Not Directly Suitable for V2A}
\label{app:suno_cover_limitation}

\begin{figure*}[t]
    \centering
    \includegraphics[width=0.6\linewidth]{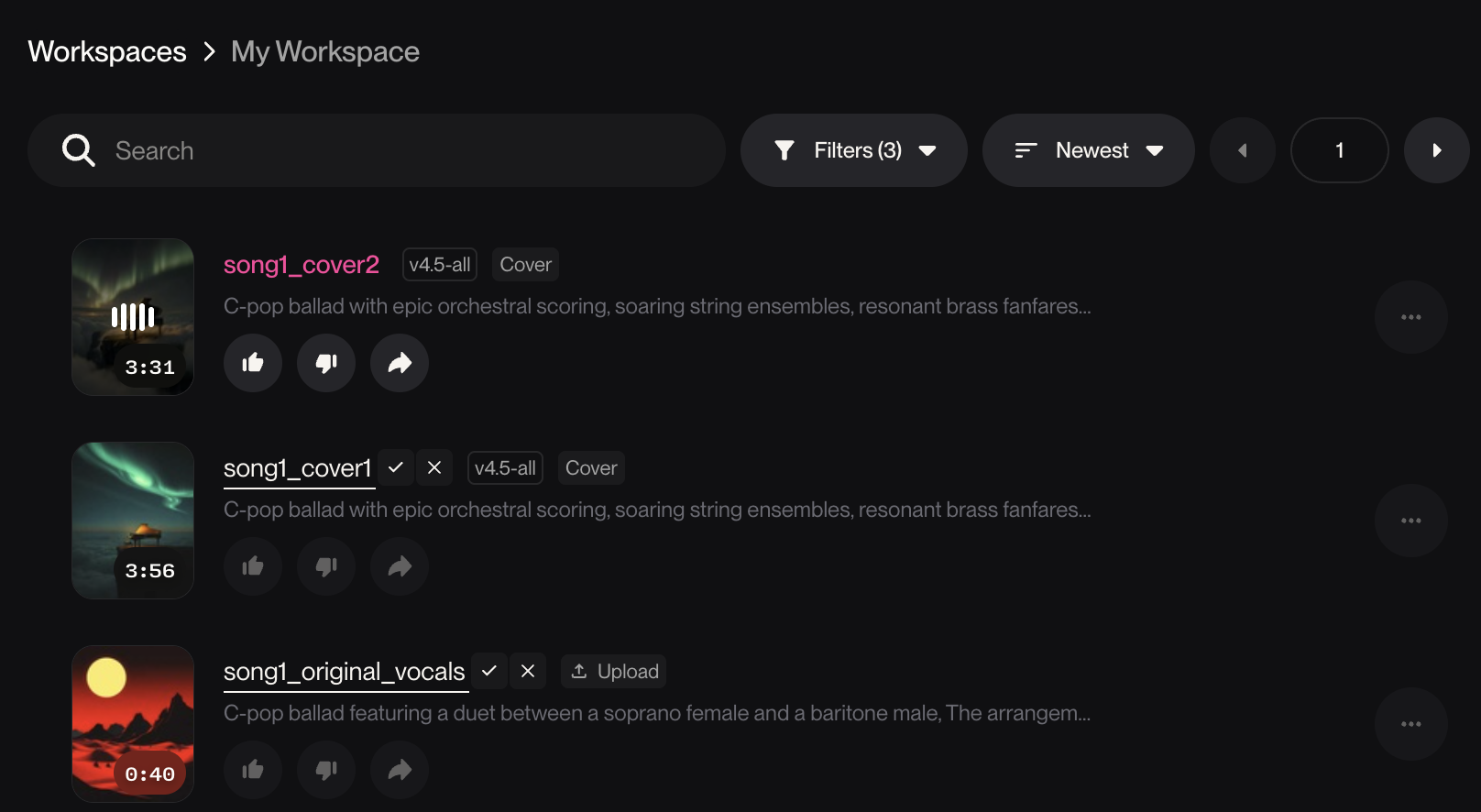}
    \caption{\textbf{General-purpose cover generation is not directly suitable for V2A accompaniment generation.} When provided with a 40-second vocal input, the system expands it into much longer full-song accompaniments (3:31 and 3:56), instead of preserving the temporal extent of the source vocal. This creates a practical gap between cover composition and directly usable vocal-aligned accompaniment generation.}
    \label{fig:cover_length_mismatch}
\end{figure*}

Figure~\ref{fig:cover_length_mismatch} illustrates a practical limitation of general-purpose cover-style composition systems when they are used for V2A generation. Given a 40-second vocal clip as input, the system produces two cover accompaniments with durations of 3:31 and 3:56, respectively. Rather than preserving the temporal extent of the source vocal, the system expands the input into a much longer full-song composition.

This behavior reveals a task mismatch between general-purpose cover generation and V2A accompaniment generation. In the V2A setting, the target accompaniment is expected to remain tightly aligned with the provided vocal, both temporally and structurally, so that it can be used directly or with minimal post-processing. By contrast, the outputs shown in Figure~\ref{fig:cover_length_mismatch} require additional manual trimming, structural rearrangement, and timeline editing before they can be paired with the original vocal as a usable accompaniment track.

\subsection{Sampling Strategy}

\begin{figure*}[h]
    \centering
    \includegraphics[width=0.7\linewidth]{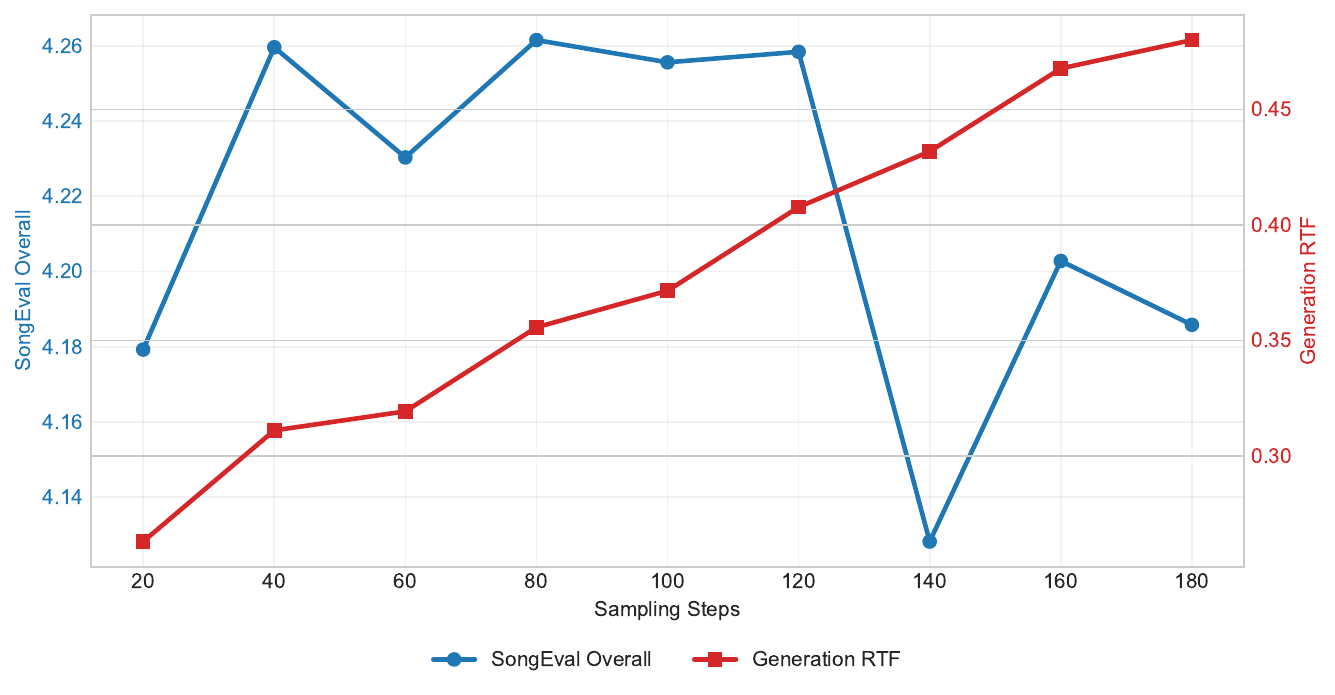}
    \caption{Effect of sampling steps on generation RTF and SongEval overall under the cosine schedule. Higher SongEval overall and lower generation RTF indicate better performance. The evaluation samples are selected from blind listening data.}
    \label{fig:steps_rtf_songeval_overall}
\end{figure*}
In the schedule ablation experiments (see Table \ref{tab:schedule_rtf_songeval_overall} and Figure \ref{fig:steps_rtf_songeval_overall}), we compared three sampling strategies: cosine, linear, and power. The results show that cosine achieved both the lowest generation RTF (0.318) and the highest SongEval overall score (4.230), demonstrating the best performance. In the steps ablation experiments, generation RTF increased approximately monotonically with the number of sampling steps, while SongEval overall did not improve continuously but peaked at around 80 steps (4.262), followed by fluctuations and even declines. This indicates that a larger number of sampling steps significantly increases generation overhead but does not necessarily lead to consistent quality gains. 

\begin{table}[t]
    \centering
    \caption{Comparison of different sampling schedules in terms of generation RTF and SongEval overall for LaDA-Band-3.0B. The evaluation samples are selected from blind listening data.}
    \label{tab:schedule_rtf_songeval_overall}
    \resizebox{0.38\textwidth}{!}{

    \begin{tabular}{lcc}
        \hline
        Schedule & Generation RTF & SongEval Overall \\
        \hline
        cosine & 0.3180 & 4.2303 \\
        linear & 0.3214 & 4.1548 \\
        power & 0.3221 & 4.2030 \\
        \hline
    \end{tabular}
    }
\end{table}

\subsection{Preliminary Model Scaling Observation}
We further conducted a preliminary scaling study by increasing the backbone size while keeping the tokenization scheme, training curriculum, and inference protocol unchanged. Under the current training setup, we did not observe a clear monotonic improvement on the automatic metrics reported in the main paper as model size increased. Nevertheless, informal listening suggests a mild qualitative trend: larger models tend to produce slightly richer instrumentation and fuller accompaniment textures. The perceptual difference is more noticeable in arrangement richness than in the current objective metrics. This suggests that, in our present regime, the benefits of scaling may not yet be fully unlocked by parameter count alone, and could depend more strongly on data quality, training budget, and long-form optimization. We leave a more systematic scaling study to future work.

\subsection{Failure Case Style Distribution on Suno70k}

\begin{figure*}[t]
  \centering
  \includegraphics[width=0.68\linewidth]{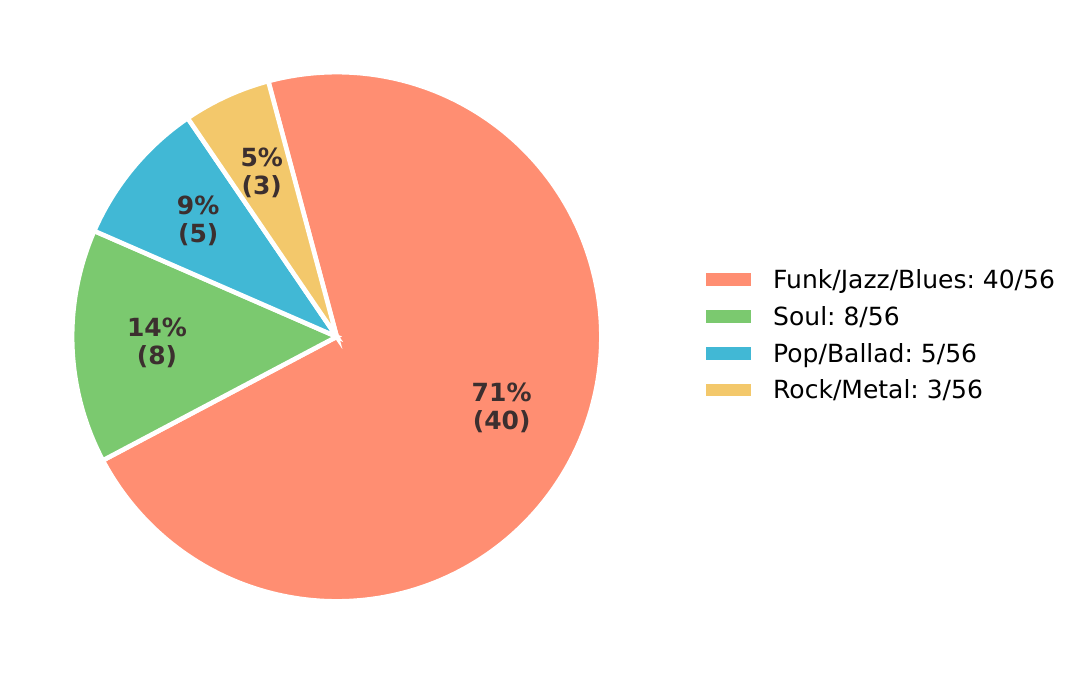}
  \caption{Style distribution of 56 negative cases on Suno70k with mean MOS below 3.0. Funk/jazz/blues accounts for the majority of the low-MOS cases.}
  \label{fig:suno70k-negative-style}
\end{figure*}

To better understand the remaining failure modes on Suno70k, we conducted an expanded manual review and collected 56 negative cases whose mean subjective MOS was below 3.0. Each case was assigned a dominant style tag by our evaluation team. As shown in Figure~\ref{fig:suno70k-negative-style}, 40 of the 56 negative cases are associated with funk/jazz/blues, 8 with soul, 5 with pop/ballad, and 3 with rock/metal.

Although this analysis is diagnostic in nature, it suggests that low-MOS cases are concentrated in groove-heavy and harmonically richer styles, especially funk/jazz/blues-related material. These styles typically require tighter rhythmic control, denser accompaniment interaction, and more nuanced harmonic coloring. By contrast, pop/ballad accounts for a relatively small portion of the negative cases, indicating that more conventional accompaniment patterns are comparatively easier for the current system to handle.

\end{document}